\documentclass[12pt]{article}%
\usepackage[nosort]{cite}
\usepackage{graphicx}
\usepackage{multicol}
\usepackage{amsfonts}
\usepackage{amssymb}
\usepackage{amsmath}
\usepackage{heck}
\usepackage{afterpage}
\usepackage{setspace}
\usepackage{verbatim}
\usepackage{color}
\usepackage{longtable}
\usepackage{float}
\usepackage{subcaption}
\usepackage{epsfig}
\usepackage{epstopdf}
\usepackage{adjustbox}
\usepackage{tikz}
\usepackage[margin=1in]{geometry}
\usepackage{titletoc}
\usepackage{hyperref}%
\providecommand{\U}[1]{\protect\rule{.1in}{.1in}}
\pdfoutput=1
\newsavebox{\mysavebox}

\hypersetup{colorlinks,citecolor=black,filecolor=black,linkcolor=black,urlcolor=black,pdftex}
\usetikzlibrary{decorations.markings}

\numberwithin{equation}{section}

\newcommand{\ba}{\begin{eqnarray}}
\newcommand{\ea}{\end{eqnarray}}

\newcommand{\be}{\begin{equation}}
\newcommand{\ee}{\end{equation}}
\tikzstyle{startstop} = [rectangle, rounded corners, minimum width=3cm, minimum height=1cm,text centered, draw=black, fill=blue!10]
\tikzstyle{startstop} = [rectangle, rounded corners, minimum width=3cm, minimum height=1cm,text centered, draw=black, fill=blue!10]
\tikzstyle{io} = [trapezium, trapezium left angle=70, trapezium right angle=110, minimum width=3cm, minimum height=1cm, text centered, draw=black, fill=blue!30]
\tikzstyle{process} = [rectangle, minimum width=3cm, minimum height=1cm, text centered, draw=black, fill=orange!30]
\tikzstyle{decision} = [diamond, minimum width=3cm, minimum height=1cm, text centered, draw=black, fill=green!30]
\tikzstyle{arrow} = [thick,->,>=stealth]
\begin{document}

\date{April 2022}

\title{Extra $W$-Boson Mass from a D3-Brane}

\institution{PENN}{\centerline{Department of Physics and Astronomy, University of Pennsylvania, Philadelphia, PA 19104, USA}}

\authors{Jonathan J. Heckman\footnote{e-mail: {\tt jheckman@sas.upenn.edu}}}

\abstract{Motivated by the recent CDF measurement of the $W$-boson mass, we study string-based
particle physics models which can accommodate this deviation from the Standard Model.
We consider an F-theory GUT in which the visible sector is realized on intersecting 7-branes,
and extra sector states arise from a probe D3-brane near an E-type Yukawa point. The D3-brane worldvolume
naturally realizes a strongly coupled sector which mixes with the Higgs.
In the limit where some extra sector states get their mass solely from Higgs vevs, this leads to a contribution to the
$\rho$ parameter which is too large, but as the D3-brane is separated from the 7-brane stack, this effect is suppressed,
leading to O(1) - O(10) TeV scale extra sector states and a correction to $\rho$ which would be in accord with the CDF result.
We also estimate the contribution to the oblique electroweak parameter $S$, and find that it is compatible with existing constraints.
This also leads to additional signatures, including diphoton resonances (as well as other diboson final states)
in the O(1) - O(10) TeV range.}

\maketitle

\enlargethispage{\baselineskip}

\setcounter{tocdepth}{2}

\newpage

\section{Introduction \label{sec:INTRO}}

Recently, CDF announced an updated measurement of the $W$-boson mass
\cite{CDF:2022hxs} which is in tension with expectations from global Standard
Model (SM) fits (see e.g. \cite{Erler:2019hds, ParticleDataGroup:2020ssz})
as well as previous measurements, e.g., from ATLAS\ \cite{ATLAS:2017rzl}
and LHCb \cite{LHCb:2021bjt}:
\begin{align}
\text{CDF\ II} &  \text{:\ }M_{W}=80,433.5 \pm9.4\text{ MeV}\\
\text{ATLAS} &  \text{: }M_{W}=80,370 \pm19.\text{ MeV}\\
\text{LHCb} &  \text{: }M_{W}=80,354 \pm32.\text{ MeV}\\
\text{SM Global Fit} &  \text{: }M_{W}=80,357 \pm 6.0\text{ MeV.}%
\end{align}
Assuming the CDF II measurement is correct, and that all other quantities such
as $c_{W}$ and $M_{Z}$ remain in accord with the SM, a shift in
the $\rho$ parameter $M_{W}^{2}/c_{W}^{2}M_{Z}^{2}$ (see \cite{Veltman:1977kh})
would need to occur at the level of $\delta\rho=2\rho_{\text{SM}}(\delta M_{W}/M_{W})$, i.e.,
a deviation of order $10^{-3}$. For recent theoretical discussions of the $W$-boson mass anomaly and
its implications for physics beyond the Standard Model,
see references \cite{Zhu:2022tpr, Fan:2022dck, Lu:2022bgw, Athron:2022qpo, Yuan:2022cpw, Strumia:2022qkt, Yang:2022gvz, deBlas:2022hdk, Addazi:2022fbj, Li:2022gwc}.

This would signal the presence of additional states
sensitive to the physics of electroweak symmetry breaking. In terms of the
higher dimension operators contributing to the precision electroweak
observables \cite{Peskin:1990zt, Peskin:1991sw} such as $\left\vert H^{\dag
}D_{\mu}H\right\vert ^{2}/(4\pi f_{\text{``nat''}})^{2}$, this suggests the
appearance of new states with a ``natural mass'' on the order of $f_{\text{``nat''}}\sim 400$
GeV. Such a low mass scale is presumably already ruled out by LHC direct searches.
However, in the limit where the new states
also have a vector-like contribution to the mass whilst still coupling to the
Higgs, one can presumably evade LHC\ direct searches and still accommodate a
non-zero shift in the $\rho$ parameter. Note that even in this limit, sizable
Yukawa couplings between the Higgs and extra states would appear to be
necessary in order to fit the observed data, indicating a preference for
strong coupling effects.

Of course, it could happen that the announced result is incorrect.

But it could also be true! Indeed, it is important to ask whether models
written down with possibly other motivations in mind might provide a natural
explanation. In this short note, we consider a particular class of models
motivated by top-down string constructions. The entire
construction can be defined in purely field theoretic terms, although the
motivation for the particular structures we present here which are quite
natural from a stringy perspective might appear \textquotedblleft
contrived\textquotedblright\ from a bottom-up perspective. Moreover, the model
comes with additional signatures which can likely be tested in the coming years.

\section{D3-Branes Near a Yukawa Point}

\begin{figure}[t!]
\begin{center}
\includegraphics[scale = 0.5, trim = {0cm 4.0cm 0cm 4.0cm}]{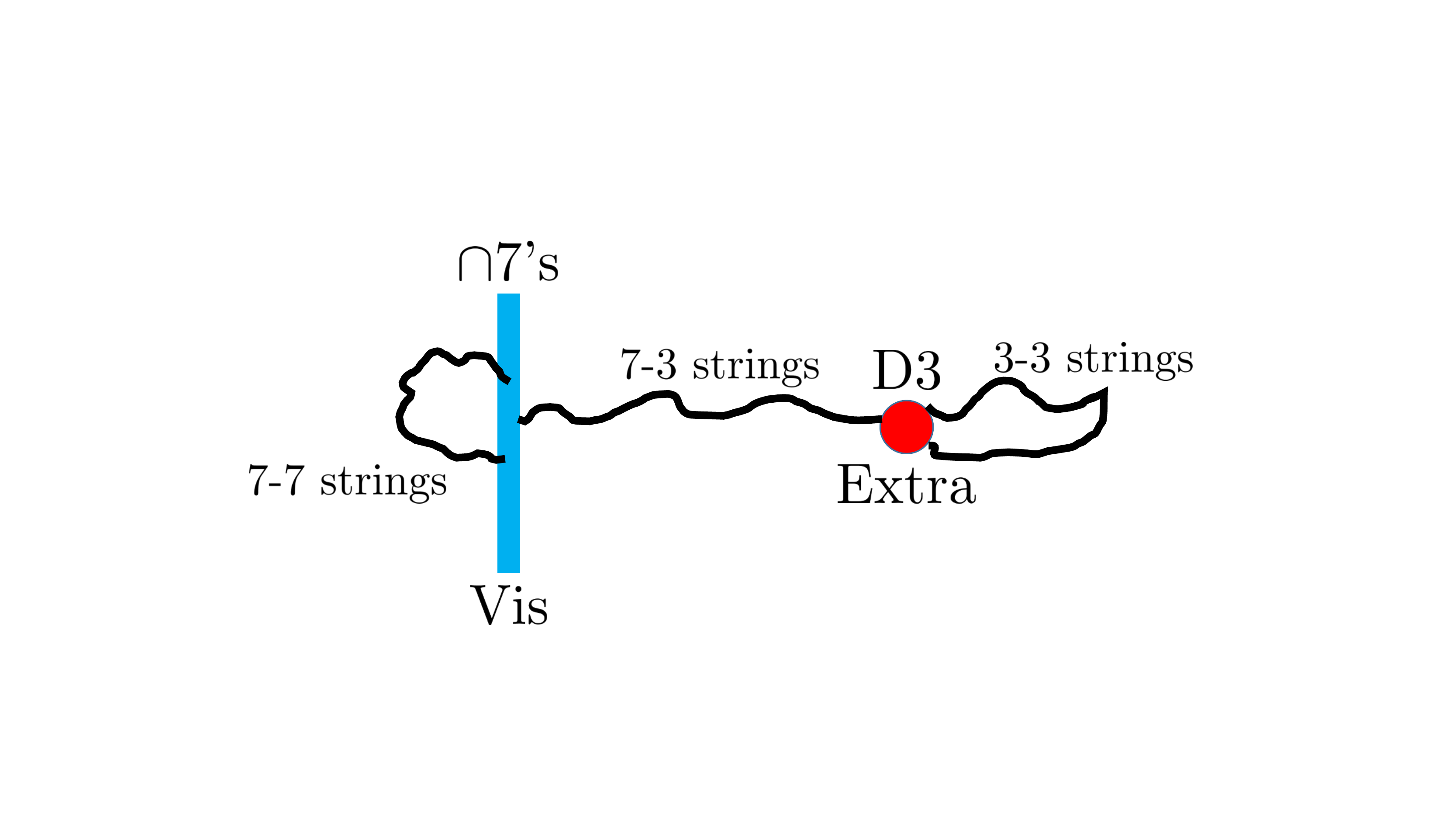}
\caption{Depiction of D3-brane probe separated from the stack of intersecting 7-brane configuration used to generate the Standard Model (i.e., visible sector). Visible sector states arise from massless $7-7^{\prime}$ strings, while the states of the D3-brane sector arise from $7-3$ and $3-3$ strings. Separating the D3-brane from the visible sector results in a mass for the $7-3$ strings and triggers a vector-like mass term for the states which mix with the Higgs fields.}
\label{fig:D3probe}
\end{center}
\end{figure}

The scenario we wish to consider is based on F-theory GUT\ models
\cite{Beasley:2008dc, Beasley:2008kw, Donagi:2008ca, Donagi:2008kj} (see
\cite{Heckman:2010bq, Weigand:2018rez} for reviews, and \cite{Cvetic:2019gnh} for a recent ``landscape scan'' of SM-like F-theory models).
These are string-based models in which the Standard Model (visible sector) is realized on a configuration of intersecting 7-branes.
A remarkable feature of this stringy particle physics scenario is that in a parametric limit where gravity decouples,
getting the correct matter content, gauge interactions and flavor structure \cite{Heckman:2008qa, Heckman:2009de, Bouchard:2009bu, Font:2013ida}
leaves little room for anything else \cite{Heckman:2009mn}. Extra sectors can still be included, but they typically arise either from other stacks of 7-branes far from the visible sector, or from mobile D3-branes which can probe the visible sector. See figure \ref{fig:D3probe} for a depiction of the different sectors of an F-theory GUT model.

We would like to understand how these models can generate a shift in the $\rho$ parameter.
In what follows, we state our model in terms of a supersymmetric theory, in part because this is the
case where we can actually calculate the quantities of interest. While there is of course no evidence for superpartners in the visible
sector, it could still happen that extra sector states are approximately supersymmetric. The main role of supersymmetry breaking
is to set an overall mass scale and effective potential for the modulus controlling the separation of the D3-brane from the
visible sector \cite{Heckman:2015kqk}.

\begin{figure}[t!]
\begin{center}
\includegraphics[scale = 0.5, trim = {0cm 1.0cm 0cm 2.0cm}]{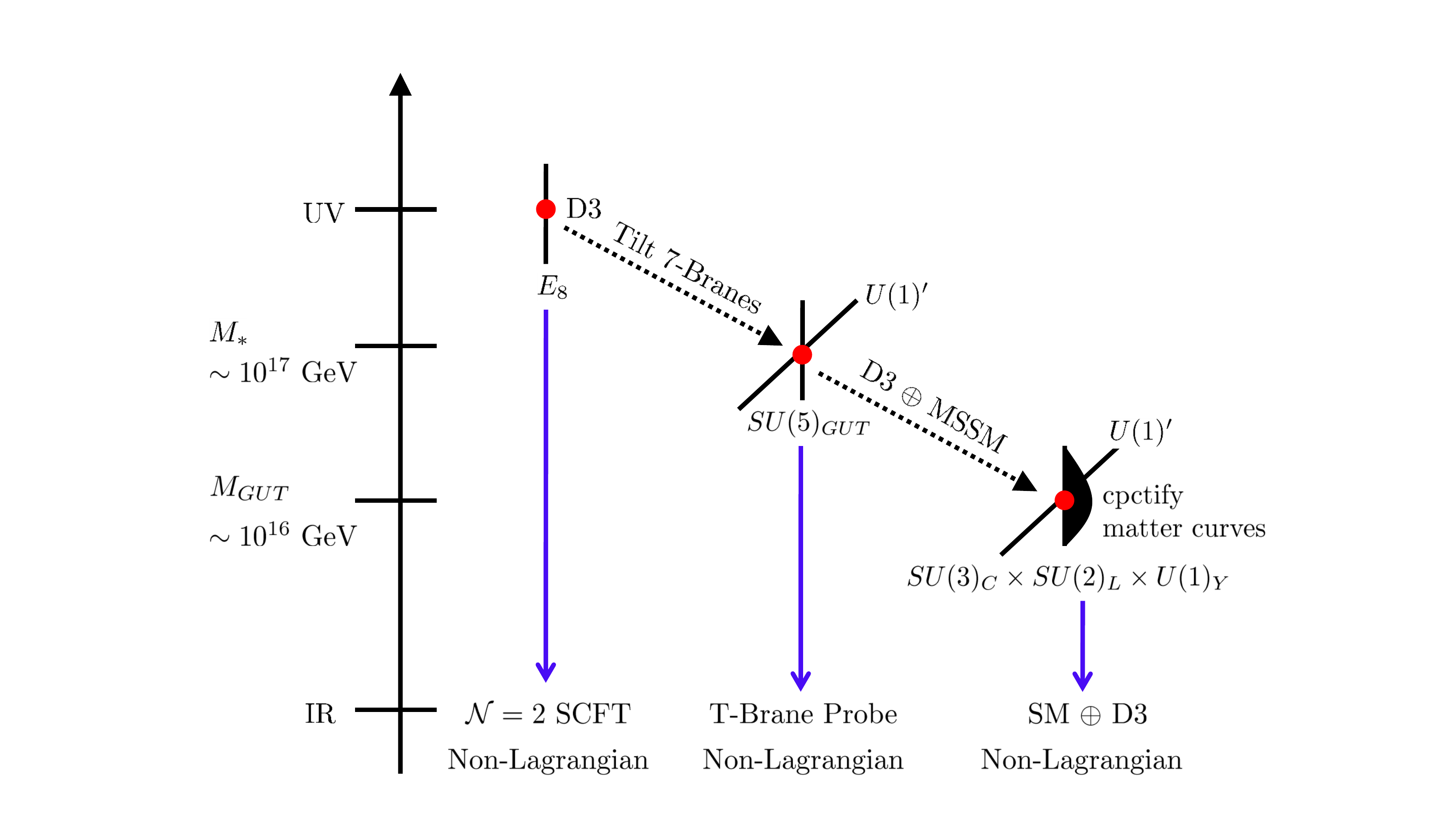}
\caption{Depiction of the different deformations of the D3-brane probe superconformal field theory (SCFT).
An F-theory GUT is generated by an intricate pattern of intersecting 7-branes.
In the limit with no tilting, the D3-brane SCFT realizes the $E_8$ Minahan-Nemschansky theory (left). Switching on 7-brane tilting with ``T-brane deformations'' (see \cite{Cecotti:2010bp, Heckman:2010qv})
generates an $\mathcal{N} = 1$ deformation to a new conformal fixed point (middle). Coupling to dynamical Standard Model fields, and in particular mixing with the Higgs sector results in another $\mathcal{N} = 1$ fixed point (right) in which contributions to the $\rho$ parameter are calculable.}
\label{fig:D3Defs}
\end{center}
\end{figure}

Now, in models with a mobile D3-brane, the same physics which generates hierarchical Yukawa couplings
also attracts D3-branes to an E-type Yukawa point \cite{Cecotti:2009zf, Marchesano:2009rz, Heckman:2010fh, Cvetic:2012ts}.
The resulting worldvolume theory is a strongly coupled supersymmetric
conformal field theory (SCFT), the formal properties and resulting phenomenology of
which were determined in a series of papers \cite{Heckman:2010fh,
Heckman:2010qv, Heckman:2011sw, Heckman:2011hu, Heckman:2011bb, Heckman:2012nt, Heckman:2012jm,
Heckman:2015kqk, DelZotto:2016fju, Balasubramanian:2020lux, Baumgart:2021ptt}. Treating, for the moment, all states of the Standard
Model as non-dynamical, the probe D3-brane theory is specified by a
particular relevant deformation of the $\mathcal{N}=2$ Minahan-Nemeschansky theory with
$E_{8}$ flavor symmetry \cite{Minahan:1996fg, Minahan:1996cj}.
The resulting CFT$_{\text{intermediate}}$ comes with a collection of operators
transforming (descending from the adjoint of $E_8$)
in various representations of the SM gauge group. In particular,
there are couplings to the Higgs doublets of the MSSM, as well as the heaviest
generation of quarks and leptons. Our interest here is in the Higgs / D3-brane
couplings, and these can be viewed as an additional relevant deformation of
CFT$_{\text{intermediate}}$ by superpotential couplings of the form:%
\begin{equation}\label{HiggsExtra}
\int d^{2}\theta\text{ } \lambda_u H_{u}\mathcal{O}_{u}+ \lambda_d H_{d}\mathcal{O}_{u}.
\end{equation}
In a weakly coupled model, these terms could be interpreted as Yukawa couplings to a
vector-like fourth generation. In the case at hand where the extra sector is strongly coupled, these
terms drive CFT$_{\text{intermediate}}$ to a new fixed point CFT$_{\text{IR}}$
in which the scaling dimension of the Higgs fields
changes to $\Delta_{u}=1+\delta_{u}$ and $\Delta_{d}=1+\delta_{d}$. This modifies
the K\"{a}hler potential for the Higgs fields, which in turn
distorts the resulting masses for the $W$- and $Z$-bosons, as well as the
spectrum of excitations in the two Higgs doublet model \cite{Heckman:2011bb}.
See figure \ref{fig:D3Defs} for a depiction of the different fixed points.

In particular, it is possible to calculate the contribution to the $\rho$
parameter from such couplings. Precisely because we are at strong coupling,
the end result is of the general form:\footnote{As explained in \cite{Heckman:2011bb, Heckman:2012jm}, to extract the $W$- and $Z$-boson masses,
we use the fact that a modification in the scaling dimension for the Higgs fields corrects its K\"ahler potential, e.g.,
$(H_u^\dag H_u + H_d^\dag H_d)^{1/\Delta}$ in a case where a custodial $SU(2)$ is retained to leading order. Indeed, for a supersymmetric theory with superfields $\Phi^i$, the corresponding kinetic term $g_{i \overline{j}} D^\mu \overline{\Phi}^{\dag \overline{j}} D_{\mu} \Phi^{i} $ then generates a shift in the $\rho$ parameter.}
\begin{equation}\label{deltarho}
\delta\rho_{\text{CFT}}\simeq f_{u}(\beta) \delta_u + f_{d}(\beta) \delta_d \simeq \delta_{\text{CFT}},
\end{equation}
where $f_u(\beta)$ and $f_d(\beta)$ are order one trigonometric functions which sum to one, and
$\tan\beta \equiv v_{u}^{1/\Delta_u}/v_{d}^{1 / \Delta_d}$ is a ratio of scaled Higgs vevs.
In the last approximation of line (\ref{deltarho}), we have made the estimate $\delta_{\mathrm{CFT}} = (\delta_u + \delta_d) / 2$ to get a ``typical'' sense of the size of the deviation. In the most phenomenologically successful scenarios such as the
\textquotedblleft Dih$_{4}^{(2)}$\textquotedblright\ model of reference
\cite{Bouchard:2009bu}, the resulting value of $\delta_{\mathrm{CFT}}$ is of order $0.01$,
although larger values can be generated in other scenarios. Returning to the
Higgs / extra sector couplings of line (\ref{HiggsExtra}), these extra sector states can be interpreted as weakly coupled states with large Yukawa couplings, with corresponding masses on the order of $f_{\text{``nat''}} \sim 400$ GeV, so we take this as a reference value in what follows.
Taken at face value, getting such a large value of $\delta_{\mathrm{CFT}}$ would result in
a contribution to $\delta \rho$ which is too large by a factor of $10$ to accommodate the CDF result, let alone the other measurements which are in accord with the Standard Model.

Of course, for various reasons, we must consider a deformation of the
CFT$_{\text{IR}}$ theory. For one thing, the presence of a large number of
additional low mass states charged under the Standard Model gauge group would have been detected
by now. The simplest, well-motivated possibility is to
assume that while the D3-brane is nearby the Standard Model stack, it is
actually somewhat removed from it, i.e., separated from it in the extra
dimensions by a characteristic scale $f_{\text{D3}}$ (see figure \ref{fig:D3probe}).
As explained in \cite{Heckman:2015kqk}, $f_{\text{D3}}$ is expected to be the same order of magnitude as the
mass scales of the extra sector. Assuming present LHC limits, we take
this to be on the order of $1$ TeV. In this case, the net contribution to
higher-dimension operators such as $\left\vert H^{\dag}D_{\mu}H\right\vert
^{2}$ will receive some additional suppression relative to the purely
conformal case. We can model such effects by introducing an effective
$\delta_{\text{eff}} \equiv \delta_{\mathrm{CFT}} \times \left(  4\pi f_{\text{``nat''}}/4\pi f_{\text{D3}}\right)
^{2}$, with $f_{\text{``nat''}}\sim 400$ GeV. All told, then, we expect
the contribution to the $\rho$ parameter from the D3-brane sector to be:%
\begin{equation}
\delta\rho_{\text{D3}}\simeq10^{-3} \times \left(
\frac{\delta_{\text{CFT}}}{0.01}\right)  \times\left(  \frac{1.3 \text{ TeV}%
}{f_{\text{D3}}}\right)  ^{2}.
\end{equation}
Surveying the list of F-theory GUT models with different spectral cover monodromy \cite{Bouchard:2009bu},
the values of $\delta_{\text{CFT}}$ for the different models
were collected in line (3.64) of \cite{Heckman:2011hu}. This results in the D3-brane mass
scales:
\begin{equation}%
\begin{tabular}
[c]{|l|l|l|l|l|l|l|}\hline
One D3-Brane & $%
%TCIMACRO{\U{2124} }%
%BeginExpansion
\mathbb{Z}
%EndExpansion
_{2}^{(1)}$ & $%
%TCIMACRO{\U{2124} }%
%BeginExpansion
\mathbb{Z}
%EndExpansion
_{2}^{(2)}$ & $%
%TCIMACRO{\U{2124} }%
%BeginExpansion
\mathbb{Z}
%EndExpansion
_{2}\times%
%TCIMACRO{\U{2124} }%
%BeginExpansion
\mathbb{Z}
%EndExpansion
_{2}$ & $S_{3}$ & Dih$_{4}^{(1)}$ & Dih$_{4}^{(2)}$\\\hline
$\delta_{\text{CFT}}$ & $0.065$ & $0$ & $0.178$ & $0.08$ & $0.145$ &
$0.01$\\\hline
$f_{\text{D3}}$ & $3.2$ TeV & X & $5.3$ TeV & $3.6$ TeV & $4.8$ TeV & $1.3$
TeV\\\hline
\end{tabular}
\ ,\label{ONED3}%
\end{equation}
where an \textquotedblleft X\textquotedblright\ indicates that we cannot fit
the shift in the $\rho$ parameter mass for any value of $f_{\text{D3}}$. We
stress that these are order of magnitude estimates since we have imperfect
knowledge of the microscopic details of the strongly coupled CFT.

One can repeat this analysis in the case of multiple D3-branes, but
beyond two D3-branes leads to a loss of asymptotic freedom and perturbative
gauge coupling unification \cite{Heckman:2011hu}. The values of $\delta_{\text{CFT}}$ for the different models
were collected in line (3.67) of \cite{Heckman:2011hu}. This results in the D3-brane mass
scales:
\begin{equation}%
\begin{tabular}
[c]{|l|l|l|l|l|l|l|}\hline
Two D3-Branes & $%
%TCIMACRO{\U{2124} }%
%BeginExpansion
\mathbb{Z}
%EndExpansion
_{2}^{(1)}$ & $%
%TCIMACRO{\U{2124} }%
%BeginExpansion
\mathbb{Z}
%EndExpansion
_{2}^{(2)}$ & $%
%TCIMACRO{\U{2124} }%
%BeginExpansion
\mathbb{Z}
%EndExpansion
_{2}\times%
%TCIMACRO{\U{2124} }%
%BeginExpansion
\mathbb{Z}
%EndExpansion
_{2}$ & $S_{3}$ & Dih$_{4}^{(1)}$ & Dih$_{4}^{(2)}$\\\hline
$\delta_{\text{CFT}}$ & $0.11$ & $0$ & $0.225$ & $0.21$ & $0.29$ &
$0.08$\\\hline
$f_{\text{D3}}$ & $4.2$ TeV & X & $6.0$ TeV & $5.8$ TeV & $6.8$ TeV & $3.6$
TeV\\\hline
\end{tabular}
\ .\label{TWOD3}%
\end{equation}

The main point is that the CFT$_\mathrm{IR}$ fixed point would generate a too
large contribution to the $\rho$ parameter, but we can decouple
the effects by introducing vector-like masses, which in some cases need to be
rather large to sufficiently suppress the resulting contributions.

It is also important to ask about the contribution from these extra
sector states to the oblique electroweak parameter $S$
\cite{Peskin:1990zt, Peskin:1991sw}.
We can estimate this contribution using the methods of \cite{Heckman:2012jm}.
In many of the D3-brane scenarios considered here, the leading order contribution is a shift of order:
\begin{equation}
\delta S \simeq \frac{\delta b_{SU(5)}}{2 \pi} \left(\frac{f_{\text{``nat''}}}{f_{\text{D3}}} \right)^2,
\end{equation}
where $\delta b_{SU(5)} \simeq 2 - 4$ is the size of the threshold correction to the unified $SU(5)$ gauge coupling. By inspection, this is typically on the order of $10^{-2}$ to $10^{-3}$, which is within the tolerance of recently updated precision electroweak fits \cite{Lu:2022bgw, Athron:2022qpo, Strumia:2022qkt, deBlas:2022hdk}.

A subtle feature of many electroweak global fits is possible mixing with other higher dimension operators.\footnote{We thank C. Kilic for comments on this point.} We expect the dominant contributions to be associated with the $S$ and $T$ (and thus implicitly $\rho$) parameters, which has been our main focus here, but of course it would be important to explore this more systematically in future work, perhaps along the lines of \cite{Strumia:2022qkt}.

\section{Further Signatures}

The requirement that we separate the D3-brane from the visible stack leads to
additional signatures. For example, in this class of scenarios the vev of a
dynamical modulus controls the separation between the D3-brane and the
visible sector. The mass of this state is roughly of the same order of
magnitude as $f_{\text{D3}}$, and will produce some additional resonances
which can in principle be detected in future LHC\ searches. This
modulus couples to the SM vector bosons via the higher dimension operator:
\begin{equation}
\frac{\delta b_{G}}{32 \pi^2} \left( \frac{s}{f_{\text{D3}}} \right) \mathrm{Tr}_{G}F_{G}^{\mu \nu} F^{G}_{\mu \nu}
\end{equation}
with $F_{\mu \nu}^{G}$ the corresponding field strength for a gauge group factor $G$.
The order one coefficient $\delta b_{G}$ in front of this dimension five operator is controlled
by the contribution from the D3-brane sector to the beta function for each
gauge group factor. In fact, precisely this mechanism was explored in
\cite{Heckman:2015kqk} as a way to generate a $750$ GeV diphoton resonance via
gluon fusion, i.e., $gg\rightarrow s\rightarrow\gamma\gamma$. To get a $5$ fb production cross section,
this required a somewhat lower value of $4 \pi f_{\text{D3}}$ of order $1$ TeV.\footnote{One might ask
whether the model of \cite{Heckman:2015kqk} was in fact compatible with the precision
electroweak constraints known at that time. In the model considered there, the
mixing from the Higgs sector was basically neglected, and this can be arranged
by tuning the $\lambda_{u} H_{u}\mathcal{O}_{u}$ and $\lambda_{d} H_{d}\mathcal{O}_{d}$
couplings to small values. One can of course reintroduce this tuning and in so doing
suppress any shift to the $\rho$ parameter.} Scaling up to the
list of $f$'s given in lines (\ref{ONED3}) and (\ref{TWOD3}), we expect a suppressed
production cross section of order $\sim 5$ fb $\times(1$ TeV$/ 4 \pi f_{\text{D3}})^{2}$.
This should be interpreted as a crude order of magnitude estimate, i.e.,
there could be a stray order one factor multiplying $f_{D3}$.\footnote{For example,
the mass of the modulus controlling the separation of the D3-brane from the visible sector will in general differ from the suppression scale $f_{\mathrm{D3}}$ entering in the dimension five operator $s \mathrm{Tr}_{G} F^2$.}

While the evidence for the $750$ GeV diphoton resonance has since collapsed,
the generic expectation that such resonant diphoton signatures will appear at
\textit{some scale} remains a feature of all of these D3-brane scenarios.
Additional signatures, as well as their rates include the corresponding
contribution from $s$ and its axionic partner $a$ include
(see \cite{Heckman:2015kqk} for further discussion):

\begin{itemize}
\item $pp\rightarrow s/a\rightarrow\gamma\gamma$

\item $pp\rightarrow s/a\rightarrow gg$

\item $pp\rightarrow s/a\rightarrow ZZ$

\item $pp\rightarrow s/a\rightarrow WW$

\item $pp\rightarrow s/a\rightarrow Z\gamma$
\end{itemize}

See, e.g., \cite{ATLAS:2021uiz} for a recent update on resonant diphoton searches. From figure 5 of \cite{ATLAS:2021uiz},
we observe that the typical production rate is below current limits in all of our scenarios, but with increased luminosity,
such scenarios may eventually be probed. For example, the current bound on a $1$ TeV diphoton scalar resonance is around $10^{-1}$ fb,
while the $\mathrm{Dih}^{(2)}_{4}$ scenario would indicate (roughly) a value of order $10^{-2}$ fb.

Of course, the overall cross section depends on various details of the model and in particular the mass of the modes $s$ and $a$. As the mass increases, the overall cross section will of course decrease further. From the perspective of a UV complete model, the mass depends on the local potential for the D3-brane probe in the F-theory model. Following \cite{Heckman:2015kqk} (see equation (2.2) of \cite{Heckman:2015kqk}), we note that non-perturbative corrections to the D3-brane superpotential induce a mass for this modulus of order $M_{GUT} (M_{IR} / M_{GUT})^{\Delta - 1}$, with $M_{IR} \sim $ TeV an infrared scale of conformal symmetry breaking, and $\Delta \sim 2$ the scaling dimension of the operator responsible for the motion of the D3-brane. This in turn leads to a mass scale in the TeV range. Supersymmetry breaking can induce an additional model dependent mass splitting between $s$ and $a$.

Another natural question is whether the presence of such a D3-brane might introduce contributions to other beyond the Standard Model
signatures such as proton decay. This question was actually considered in reference \cite{Heckman:2011sw}. It was found there that the coupling to a conformal extra sector has a large conformal suppression in the resulting higher-dimension operators. An additional comment here is that the D3-brane probe theory enjoys approximate global $U(1)$ symmetries which further suppress such contributions to proton decay.

\section{Conclusions}

The reported $W$-boson mass measurement by CDF is quite
exciting. Though it will need to be confirmed, it is already worth asking how well
various scenarios for physics beyond the Standard Model can accommodate this measurement.
While there are clearly many possible models which can do the job
(see e.g., \cite{Zhu:2022tpr, Fan:2022dck, Lu:2022bgw, Athron:2022qpo, Yuan:2022cpw, Strumia:2022qkt, Yang:2022gvz, deBlas:2022hdk, Addazi:2022fbj, Li:2022gwc}),
in this note we have explained how top-down motivated F-theory GUT models
with an extra sector generated by a probe D3-brane provide a
natural candidate scenario which might otherwise appear
\textquotedblleft contrived\textquotedblright\ from purely bottom-up
considerations. Thankfully, much of the required formal analysis of Higgs /
D3-brane sector mixing was already performed in previous work. For suitable choices of mass scales and parameters,
this class of models can indeed produce shifts in the $\rho$ parameter of the correct size, whilst remaining compatible
with constraints on the $S$-parameter and direct searches at the LHC.
The model also predicts additional signatures, especially with respect to vector boson
production. This holds out the prospect of vaulting the gap from the electroweak scale to the Planck
scale. We anticipate that the coming years will provide additional clarity.

\section*{Acknowledgements}

We thank C. Kilic for helpful correspondence and comments on an earlier draft,
and J. Erler and E. Lipeles for helpful correspondence.
The work of JJH is supported by DOE (HEP) Award DE-SC0013528.

%%%%%%%%%%%%%%%%%%%%%%%%%%%%%%%%%%%%%%%%%%%%%%%%%%%%%%%%%%%%%%%%%%%%%%%%%%%%

%\newpage

\bibliographystyle{utphys}
\bibliography{D3WBosons}

\end{document}